\documentclass[11pt,preprint]{aastex}
\usepackage{natbib}
\usepackage{graphicx}
\begin{document}
\title{X-ray Outflows in the Swift Burst Alert Detected Seyfert 1s}
\author{Lisa M. Winter\altaffilmark{1,2}
\altaffiltext{1}{Center for Astrophysics and Space Astronomy, University of Colorado, Boulder, CO}
\altaffiltext{2}{Hubble Fellow}}

\begin{abstract}
Previous surveys of outflows in low-redshift active galactic nuclei (AGN) have relied on the analysis of sources selected primarily for their optical/X-ray brightness, and are therefore biased.  Towards determining the outflow properties of local AGN, we detect warm absorption signatures of \ion{O}{7} and \ion{O}{8} absorption edges in the available Suzaku/XMM-Newton CCD spectra of  an unbiased sample of 44 Seyfert 1--1.5 sources selected in the very hard X-rays (14--195\,keV) with the Swift Burst Alert Telescope.  From our analysis, we find that \ion{O}{7} and \ion{O}{8} absorption edges are present in 41\% of the sample.  This fraction is dependent on luminosity, with outflow detections in 60\% of low luminosity and 30\% of high luminosity sources.  However, grating spectroscopy of the highest luminosity sources reveals that $\sim 80$\% of these sources have ionized absorbers, but that the ionization states are higher/lower than produces the \ion{O}{7} and \ion{O}{8} edges.  This suggests that ionized absorption may be present in all local Seyfert 1s.
\end{abstract}

\section{Introduction}
Mass outflows from active galactic nucleus sources (AGNs) are believed to affect their surrounding galaxy and the intergalactic environment in many ways.  The powerful winds from AGN may play an active role in shaping the host galaxy evolution by blowing out the necessary fuel for star formation, effectively quenching it.  Further, the ejected gas enriches the galaxy's local environment as well as the intergalactic medium with metals produced from supernovae.  While it is believed that outflows are ubiquitous in AGN, signatures of these winds are only seen in $\approx$ 50--60\% of Seyfert (Sy)\,1--1.5 sources in the X-ray and UV bands 
\citep{1997MNRAS.286..513R,1998ApJS..114...73G,1999ApJ...516..750C}.  This fraction is of fundamental importance, since it represents the solid angle occupied by the outflow.  


The previous outflow studies are not complete -- selecting bright sources in the optical/soft X-ray bands. Therefore, they may be biased towards selecting warm absorbers. To determine the true fraction of AGN with outflows/warm absorbers, it is necessary to search for absorption signatures in an unbiased sample.  In this letter, we detail our study identifying outflows in the sources selected in the very hard X-ray band with Swift's Burst Alert Telescope (BAT).  Due to the selection in the 14--195\,keV band, the Swift sample is unbiased towards all but the heaviest line-of-sight absorption ($N_H > 10^{24}$\,atoms\,cm$^{-2}$) for sources with a BAT flux of $\ga 2 \times 10^{-11}$\,ergs\,s$^{-1}$\,cm$^{-2}$.


\section{The Swift Sample}
 The sample of 153 AGNs in the Swift BAT nine-month catalog  \citep{2008ApJ...681..113T} are the most well-studied and brightest hard X-ray sources.  Details of their 0.3--10\,keV X-ray spectra \citep{2009ApJ...690.1322W}, optical spectra \citep{2010ApJ...710..503W}, and infrared spectra from Spitzer \citep{2010ApJ...716.1151W} are well documented in the literature.  Additionally, many of these sources are well-known AGNs with high signal-to-noise archived data available from current X-ray observatories (i.e.,~Suzaku, XMM-Newton, Chandra).

We selected all of the optically classified Seyfert 1--1.5 sources in the nine-month BAT catalog  with a high Galactic latitude ($|b| > 15^{\circ}$).  This sample includes 51 sources, of which 44 have archived spectra available.  The typically low intrinsic column densities (N$_{\rm H} \le 10^{21}$\,cm$^{-2}$) of Seyfert 1s make them the best sources for searching for outflow signatures.  This is due to the fact that X-ray outflow signatures are found primarily in the soft band, where more absorbed sources have lower observed luminosities.

\section{X-ray Observations}
In the X-ray band, outflows are detected primarily in the soft 0.3--2\,keV band through emission and/or absorption features from oxygen (e.g.,~0.65 keV \ion{O}{8} Ly$\alpha$, 0.77 keV \ion{O}{8} Ly$\beta$), magnesium (e.g.,~1.35\,keV \ion{Mg}{11}), neon (e.g.,~1.02\,keV \ion{Ne}{10} Ly$\alpha$), and iron  (e.g.,~0.87 keV \ion{Fe}{18}, 0.92 keV \ion{Fe}{19}, 0.96 keV \ion{Fe}{20}).  However, detection of these lines typically requires high signal-to-noise observations with the grating spectrometers on XMM-Newton or Chandra.  To obtain high signal-to-noise grating spectra, long exposures are required of the order of $\ga 100$\,ks for the Swift sources.  Therefore, it is more feasible to detect X-ray outflows in a larger sample of sources through CCD data.  In the CCD data, absorption edges from \ion{O}{7} (0.73\,keV) and \ion{O}{8} (0.87\,keV) indicate the presence of outflows.

High signal-to-noise archived X-ray CCD data from Suzaku and XMM-Newton are available for many of our sources.  While the \ion{O}{7} and \ion{O}{8} edges are detectable in data of moderate quality (i.e., $
\sim 10$\,ks with XMM-Newton for the Swift sources as in \citealt{2008ApJ...674..686W}), we chose to use predominantly higher exposure time observations of $\ga 40$\,ks (though, exposure times range from 2--275\,ks with 34/44 having exposure times $> 40$\,ks).  The Suzaku X-ray observatory was our first choice for X-ray observations since simultaneous data is available from the 0.3--12\,keV (with the XIS detectors) and $\sim 15$--50\,keV band (with the HXD PIN detector).  This allows us to better constrain the underlying continuum and provides overlapping energy coverage between the Suzaku spectra and the 22-month averaged 14--195\,keV Swift BAT spectra \citep{2010ApJS..186..378T}.  Properly constraining the continuum emission is essential in order to accurately determine limits on the optical depth of  \ion{O}{7} and \ion{O}{8} absorption edges.  Suzaku data were available for 27/51 sources.  Where multiple Suzaku observations exist, we analyzed the longest observation.

For sources without a Suzaku follow-up, we analyzed archived XMM-Newton EPIC observations.  An additional 17 sources have high signal-to-noise XMM-Newton observations without a Suzaku follow-up.  In total, we analyzed the spectra of 44/51 (80\%) of the Seyfert 1--1.5 sources in the Swift BAT nine-month catalog.  The sources which do not have available high quality observations represent lower luminosity sources which were identified as AGN for the first time by Swift.

Full details of the observations and data are presented in Winter {\it et al.} (in prep).  To summarize the processing of the observations, for the Suzaku XIS (XIS0, XIS1, XIS3, and XIS2, where available -- in the 0.3--12\,keV band) and XMM-Newton EPIC pn (in the 0.3--10\,keV band) data we extracted the spectra from circular regions centered on the source of $\sim 60$--200\arcsec.  Background regions were extracted in regions near the source but free of additional sources.  The Suzaku spectra were extracted using the {\tt ftool} {\tt xselect}, while the pn spectra were extracted with SAS using the standard filtering for the pn as detailed in the SAS ABC Guide.  Response and ancillary response matrices were generated using the Suzaku {\tt ftools} {\tt xisrmfgen} and {\tt xisarfgen} or SAS commands {\tt rmfgen} and {\tt arfgen}.  The spectra were then grouped and binned to 20\,counts\,per\,bin using {\tt grppha}.

The Suzaku HXD PIN ($\sim 15$--50\,keV; \citealt{2007PASJ...59S..35T}) spectra were extracted using {\tt xselect} from the processed cleaned event files. The tuned PIN background file specific to each observation was used to extract a background spectrum.  The spectra were corrected for dead time and the CXB contribution was added to the background spectrum as specified in the Suzaku ABC Guide.  The corresponding response file from the Suzaku CALDB was used for each observation.  The PIN spectra and responses were grouped using {\tt grppha} and binned to a signal-to-noise of 3--4$\sigma$.  Finally, we also utilized the Swift BAT 22-month averaged spectra and diagonal response file from \citet{2010ApJS..186..378T}.  These spectra consist of 8 energy channels in the BAT band of 14--195\,keV. 

\section{Spectral Analysis}  
We simultaneously fit the Suzaku XIS and PIN or XMM-Newton pn spectra with the Swift BAT spectra to obtain joint fits in the 0.3--195\,keV band.  We used the {\tt XSPEC} \citep{ADASS_96_A} spectral fitting package v12.  For the sources with Suzaku spectra, we added a constant model to allow for differences in the flux scaling between the spectra, normalizing to the XIS1 spectrum.  The PIN spectrum was fixed to the value of 1.16, as detailed in the Suzaku ABC Guide, while the constant factor was allowed to vary for the BAT spectrum and the additional XIS spectra.  For sources with XMM-Newton spectra, we allowed the constant factor to vary relative to the pn level.

For all of the sources, we fit the broad-band X-ray spectra with a base cutoff power law model.  Seyfert direct emission is assumed to be produced from inverse Compton scattering in the corona surrounding an accretion disk -- which is well represented by a power law with a cutoff at high energy.  We included neutral absorption from the Galaxy with a {\tt tbabs} model, fixing the Galactic column density to the \citet{1990ARAA..28..215D} value.  Where the addition of intrinsic neutral absorption improves the fit ($\Delta\chi^2 \ga 6.64$ on adding the model, corresponding to $P = 0.01$ for the addition of one degree of freedom), we added intrinsic emission with a {\tt ztbabs} model.  Where a soft excess was evident, we included a blackbody model to account for the soft excess. 
To better constrain the continuum emission, we added a Gaussian model ({\tt zgauss}) to account for neutral Fe K$\alpha$ emission and a {\tt pexrav} model to account for reflected emission from the accretion disk.  Full details of the modeling, including the best-fit parameters (in particular, L$_{14-195\,{\rm keV}}$ and N$_{\rm H}$), are in Winter et al. (in prep).  

With a baseline model for the continuum, we searched the spectra for signatures of outflows through the presence of \ion{O}{7} and \ion{O}{8} edges.  We used two {\tt zedge} models, one to account for each of the two absorption edges.  The {\tt zedge} model has three parameters, the energy of the edge, redshift, and optical depth.  We initially allowed the edge energy and optical depth to vary around the laboratory measured values of the edges (0.73 and 0.87\,keV, respectively), fixing the redshift to the source redshift.  Where the energies were not well-constrained (i.e., the fitted energies were not consistent with the absorption edge energies), we fixed the energies to the laboratory value.  In Table~\ref{tbl-warmabs}, we record the results of this analysis, including the $\Delta\chi^2$ upon adding the absorption edges to the baseline model, energies, and optical depths for the fitted edges.  We classify the source as exhibiting an outflow where an improvement in $\chi^2$ corresponds to $\Delta\chi^2 \ga 13.39$ (a probability of $P = 0.01$ for four additional degrees of freedom).  Additionally, we include NGC 3516, whose spectrum was not well-fit by our base model, but clearly has an outflow present (see \citealt{2008PASJ...60S.277M} for a complete analysis of the absorption in the Suzaku observation of NGC 3516).

\section{Results}
We classify 41\% (18/44) of Swift BAT Sy 1--1.5 sources as exhibiting the outflow signatures probed through detection of \ion{O}{7} and \ion{O}{8} edges.  Of the outflow sources, half are Sy 1--1.2s and half are Sy 1.5s.  The range of optical depths for \ion{O}{7} is $\tau \approx 0.07 - 2.54$ and for \ion{O}{8} is $\tau \approx 0 - 1.66$.  For non-outflow sources, we find the range of optical depths for \ion{O}{7} is $\tau \approx 0 - 0.29$ and for \ion{O}{8} is $\tau \approx 0 - 0.26$.  In this section, we describe the properties of the outflow versus non-outflow sources.

In Figure~1, we plot the optical depth of \ion{O}{7} and \ion{O}{8} versus the Swift BAT luminosity (14--195\,keV) for the outflow (circle) and non-outflow (triangle) sources in our sample.  Throughout this section, we use the Swift BAT luminosity because it is measured at high energies and is thus less affected by obscuration.  Clearly, the outflow sources have the highest measurements for optical depth, on average.  Additionally, we find that the measured optical depth in both \ion{O}{7} and \ion{O}{8} is low for the highest luminosity ($L_{14-195 {\rm keV}} \ga 5 \times 10^{43}$\,ergs\,s$^{-1}$) sources overall.  We plot the Sy 1--1.2 sources with open and the Sy 1.5 sources with filled symbols.  Another trend which we find is that the highest luminosity sources tend to be Sy 1--1.2 sources.  Only one source out of eleven with $L_{14-195 {\rm keV}} \ga 10^{44}$\,ergs\,s$^{-1}$ is a Sy 1.5.  Finally, the four sources with the strongest outflows (NGC 3516, NGC 4151, Mrk 6, and NGC 526A) have intermediate luminosities and are all Sy 1.5s.

Additionally in Figure~1, we plot the relationship between optical depth and accretion rate for our sample.  As an estimate of accretion rate, we use L$_{14-195 {\rm keV}}$/L$_{\rm Edd}$, where L$_{\rm Edd} = 1.38 \times 10^{38} {\rm M}/{\rm M}_{\sun}$\,ergs\,s$^{-1}$.  The black hole mass estimates are from stellar bulge K-band photometry in \citet{2009ApJ...690.1322W}, which were shown to be well-correlated with reverberation mapping based mass estimates (see  \citealt{2010ApJ...710..503W}).  There is no correlation between accretion rate and optical depth.  However, the sources with the strongest outflows have intermediate values of accretion rates and luminosities.    The highest accretion rate sources are Sy 1--1.2 sources. 

Table~\ref{tbl-compare} includes average, standard deviation, and the results of Kolmogrov-Smirnov (K-S) statistical tests for measured parameters of our outflow and non-outflow samples.  Comparing the average values of both luminosity and accretion rate, we find that the sources with outflows have lower luminosities and accretion rates than the sources without outflows. Also, we find that the non-outflow sources have more massive black holes. There is also a difference between the outflow and non-outflow sources in the measured neutral hydrogen column density.  Assuming N$_{\rm H} = 10^{19}$\,cm$^{-2}$ for sources with no significant $N_{H}$ measured, we find that sources with detected outflows have larger neutral column densities.

Results of the K-S tests show that the probability of mass and accretion rates of the outflow and non-outflow sources being drawn from different populations is low.  Therefore, the main statistical difference between the outflow and non-outflow sources is the luminosity and secondarily N$_{\rm H}$.

In Figure~2, we plot the fraction of outflows detected in binned N$_{\rm H}$, BAT luminosity, black hole mass, and accretion rate.  We find that there are few outflows ($< 30$\%) detected in sources with low column densities (N$_{\rm H} \la 10^{20}$\,cm$^{-2}$).  Conversely, outflows are detected in $\sim 63$\% of sources with N$_{\rm H} \ga 10^{20}$\,cm$^{-2}$.  Outflow detection rates are low at the highest luminosities ($\sim 20$\% for L$_{14-195 {\rm keV}} \ga 10^{44}$\,ergs\,s$^{-1}$), masses ($\sim 27$\% for M/M$_{\sun} \ga 3 \times 10^{8}$), and accretion rates ($\sim 27$\% for L$_{14-195 {\rm keV}}$/L$_{\rm Edd} \ga 0.003$).  Therefore, even though the total distribution of black hole mass and accretion rate between outflow and non-outflow sources is not statistically different, the binned values show fewer outflows at the highest mass/accretion rate.  At the lowest masses/accretion rates/luminosities, we find outflows in $\sim 50$--$60$\% of the sources -- approximately twice the detection rate for the highest mass/accretion rate/luminosity sources.



\section{Discussion}
We have categorized sources in our study as outflow or non-outflow sources using the significance of \ion{O}{7} and \ion{O}{8} absorption edges to a base continuum model (an absorbed power-law with soft excess and reflection).  This ensures that we are selecting the sources with the highest optical depth in \ion{O}{7} and \ion{O}{8} as outflow sources (see Figure~1).  Our sample includes 44 Sy 1--1.5 sources detected in the very hard X-rays with the Swift BAT.  This is the first study of the occurrence of outflows in an unbiased sample of AGN.


We find an outflow detection rate of 41\%.  In a similar study detecting outflows through \ion{O}{7} and \ion{O}{8} edges in ASCA spectra of 24 type 1 AGN, a detection rate of 50\% is found \citep{1997MNRAS.286..513R}.  Of these 24 AGN, 18 are also found in our survey.  We find the same categorization as an outflow/non-outflow source for 14/18 sources.  The remaining four (IC 4329A, NGC 5548, 3C 382, and 3C 390.3) are categorized as having outflows in  \citet{1997MNRAS.286..513R} and non-outflow sources in our study.  The upper limits on the optical depths for IC 4329A and NGC 5548 are within the range of values for sources with detected outflows, though the $\Delta\chi^2$ values are low.  Further, Chandra X-ray grating observations of these sources show ionized absorption is present \citep{2005AA...432..453S,2005A&A...434..569S}.  The residuals to our base model show evidence of potential warm absorber signatures in the spectra of IC 4329A and NGC 5548, but at lower energies ($< 0.7$\,keV) than probed with the \ion{O}{7} and \ion{O}{8} edges.  These features are likely associated with absorption in \ion{N}{7} and \ion{O}{7} from $\sim 0.50-0.55$\,keV.

Grating spectroscopy with XMM-Newton of 3C 382 revealed a highly ionized ($\log \xi = 2.69$\,erg\,cm\,s$^{-1}$) absorber with a column of $10^{22}$\,cm$^{-2}$ \citep{2010MNRAS.401L..10T}.  Finally, ionized absorption is detected at higher energies than probed in our study (6.6\,keV) in XMM-Newton observations of 3C 390.3, indicating an even higher ionization parameter of $\log \xi = 3.43$\,erg\,cm\,s$^{-1}$ \citep{2009ApJ...700.1473S}.  Therefore, we miss the detection of outflows in these luminous sources because they are more highly ionized.  It is unclear, however, why \ion{O}{7} and \ion{O}{8} edges were detected in the ASCA data, but not with the Suzaku/XMM-Newton CCD data.


The main question that our results raise is why there are less outflow detections in the most luminous AGN. The most luminous sources in our sample, those with $\log L_{14-195\,{\rm keV}} \ga 44.5$, are, from most to least luminous, 4C +74.26, MR 2251--178, 3C 390.3, 3C 382, 1H 0419--577, and Mrk 926.  The four highest luminosity sources are radio-loud and have very broad optical emission line profiles, including double-peaked hydrogen Balmer lines. Very highly ionized ($\log \xi \ga 3$) gas is detected in each of these sources (grating spectroscopy reveals highly ionized gas in 4C +74.26 and MR 2251--178, see \citealt{2004IAUS..222...41K}).   Of the additional luminous sources, XMM-Newton grating spectroscopy of 1H 0419--577 reveals ionized gas with a low ionization parameter ($\log \xi \sim 1.3$) and multiple observations show that the ionized (as well as neutral) absorption is variable \citep{2004ApJ...616..696P}.  Since high resolution observations of Mrk 926 do not exist, it is unclear whether ionized absorbers are present in this source.  However, it is clear that the fraction of ionized absorbers is high even in the highest luminosity sources (at least 5/6), despite the fact that \ion{O}{7} and \ion{O}{8} absorption edges are only detected in 2/6 of these sources.  Similarly, 3/4 sources with $\log L_{14-195\,{\rm keV}} < 42.5$ have absorbers detected (two are outflow sources in this paper, NGC 7213 has detected O\,VII and O\,VIII emission lines in HETG spectra
\citep{2005Ap&SS.300...81S}, and UGC 6728 has no high resolution spectra available).

The next step in understanding the outflow properties of local Seyfert 1s is to identify the more detailed properties of the ionized gas (i.e., ionization parameter, column density, velocity).  Our study of outflow detections from \ion{O}{7} and \ion{O}{8} absorption edges probes only a narrow range of ionization states.  As we found for the highest luminosity sources, we are missing the detection of more or less ionized gas and therefore underestimating the fraction of sources with outflows.  Based on a literature search of the highest luminosity sources, we find that the fraction of outflows is $\ga 80$\%, while the fraction of sources with \ion{O}{7} and \ion{O}{8} absorption edges is from 30--60\% (with low detection rates at high luminosity and high detection rates at low luminosity).  This suggests that outflows may be present in all local AGNs.  As future work, we will extend our outflow study by analyzing archived grating spectroscopy to determine the ionization state, column density, and velocity of the warm absorbers present in the Swift BAT sample.

\acknowledgements
LMW thanks Tatiana Taylor for assistance with data reduction.  She also acknowledges support through NASA grant HST-HF-51263.01-A, through a Hubble Fellowship from the Space Telescope Science Institute.  
{\it Facilities:} \facility{Swift()}, \facility{XMM()}, \facility{Suzaku()}


\begin{thebibliography}{20}
\expandafter\ifx\csname natexlab\endcsname\relax\def\natexlab#1{#1}\fi

\bibitem[{{Arnaud}(1996)}]{ADASS_96_A}
{Arnaud}, K. 1996, Astronomical Data Analysis Software and Systems V, 101, 5

\bibitem[{{Crenshaw} {et~al.}(1999){Crenshaw}, {Kraemer}, {Boggess}, {Maran},
  {Mushotzky}, \& {Wu}}]{1999ApJ...516..750C}
{Crenshaw}, D.~M., {Kraemer}, S.~B., {Boggess}, A., {Maran}, S.~P.,
  {Mushotzky}, R.~F., \& {Wu}, C.-C. 1999, \apj, 516, 750

\bibitem[{{Dickey} \& {Lockman}(1990)}]{1990ARAA..28..215D}
{Dickey}, J.~M., \& {Lockman}, F.~J. 1990, \araa, 28, 215

\bibitem[{{George} {et~al.}(1998){George}, {Turner}, {Netzer}, {Nandra},
  {Mushotzky}, \& {Yaqoob}}]{1998ApJS..114...73G}
{George}, I.~M., {Turner}, T.~J., {Netzer}, H., {Nandra}, K., {Mushotzky},
  R.~F., \& {Yaqoob}, T. 1998, \apjs, 114, 73

\bibitem[{{Kaspi}(2004)}]{2004IAUS..222...41K}
{Kaspi}, S. 2004, in IAU Symposium, Vol. 222, The Interplay Among Black Holes,
  Stars and ISM in Galactic Nuclei, ed. {T.~Storchi-Bergmann, L.~C.~Ho, \&
  H.~R.~Schmitt}, 41--44

\bibitem[{{Markowitz} {et~al.}(2008){Markowitz}, {Reeves}, {Miniutti},
  {Serlemitsos}, {Kunieda}, {Yaqoob}, {Fabian}, {Fukazawa}, {Mushotzky},
  {Okajima}, {Gallo}, {Awaki}, \& {Griffiths}}]{2008PASJ...60S.277M}
{Markowitz}, A. {et~al.} 2008, \pasj, 60, 277

\bibitem[{{Pounds} {et~al.}(2004){Pounds}, {Reeves}, {Page}, \&
  {O'Brien}}]{2004ApJ...616..696P}
{Pounds}, K.~A., {Reeves}, J.~N., {Page}, K.~L., \& {O'Brien}, P.~T. 2004,
  \apj, 616, 696

\bibitem[{{Reynolds}(1997)}]{1997MNRAS.286..513R}
{Reynolds}, C.~S. 1997, \mnras, 286, 513

\bibitem[{{Sambruna} {et~al.}(2009){Sambruna}, {Reeves}, {Braito}, {Lewis},
  {Eracleous}, {Gliozzi}, {Tavecchio}, {Ballantyne}, {Ogle}, {Barth}, \&
  {Tueller}}]{2009ApJ...700.1473S}
{Sambruna}, R.~M. {et~al.} 2009, \apj, 700, 1473

\bibitem[{{Starling} {et~al.}(2005){Starling}, {Page}, {Branduardi-Raymont},
  {Breeveld}, {Soria}, \& {Wu}}]{2005Ap&SS.300...81S}
{Starling}, R.~L.~C., {Page}, M.~J., {Branduardi-Raymont}, G., {Breeveld},
  A.~A., {Soria}, R., \& {Wu}, K. 2005, \apss, 300, 81

\bibitem[{{Steenbrugge} {et~al.}(2005{\natexlab{a}}){Steenbrugge}, {Kaastra},
  {Crenshaw}, {Kraemer}, {Arav}, {George}, {Liedahl}, {van der Meer},
  {Paerels}, {Turner}, \& {Yaqoob}}]{2005A&A...434..569S}
{Steenbrugge}, K.~C. {et~al.} 2005{\natexlab{a}}, \aap, 434, 569

\bibitem[{{Steenbrugge} {et~al.}(2005{\natexlab{b}}){Steenbrugge}, {Kaastra},
  {Sako}, {Branduardi-Raymont}, {Behar}, {Paerels}, {Blustin}, \&
  {Kahn}}]{2005AA...432..453S}
{Steenbrugge}, K.~C., {Kaastra}, J.~S., {Sako}, M., {Branduardi-Raymont}, G.,
  {Behar}, E., {Paerels}, F.~B.~S., {Blustin}, A.~J., \& {Kahn}, S.~M.
  2005{\natexlab{b}}, \aap, 432, 453

\bibitem[{{Takahashi} {et~al.}(2007){Takahashi}, {Abe}, {Endo}, {Endo}, {Ezoe},
  {Fukazawa}, {Hamaya}, {Hirakuri}, {Hong}, {Horii}, {Inoue}, {Isobe}, {Itoh},
  {Iyomoto}, {Kamae}, {Kasama}, {Kataoka}, {Kato}, {Kawaharada}, {Kawano},
  {Kawashima}, {Kawasoe}, {Kishishita}, {Kitaguchi}, {Kobayashi}, {Kokubun},
  {Kotoku}, {Kouda}, {Kubota}, {Kuroda}, {Madejski}, {Makishima}, {Masukawa},
  {Matsumoto}, {Mitani}, {Miyawaki}, {Mizuno}, {Mori}, {Mori}, {Murashima},
  {Murakami}, {Nakazawa}, {Niko}, {Nomachi}, {Okada}, {Ohno}, {Oonuki}, {Ota},
  {Ozawa}, {Sato}, {Shinoda}, {Sugiho}, {Suzuki}, {Taguchi}, {Takahashi},
  {Takahashi}, {Takeda}, {Tamura}, {Tamura}, {Tanaka}, {Tanihata}, {Tashiro},
  {Terada}, {Tominaga}, {Uchiyama}, {Watanabe}, {Yamaoka}, {Yanagida}, \&
  {Yonetoku}}]{2007PASJ...59S..35T}
{Takahashi}, T. {et~al.} 2007, \pasj, 59, 35

\bibitem[{{Torresi} {et~al.}(2010){Torresi}, {Grandi}, {Longinotti},
  {Guainazzi}, {Palumbo}, {Tombesi}, \& {Nucita}}]{2010MNRAS.401L..10T}
{Torresi}, E., {Grandi}, P., {Longinotti}, A.~L., {Guainazzi}, M., {Palumbo},
  G.~G.~C., {Tombesi}, F., \& {Nucita}, A. 2010, \mnras, 401, L10

\bibitem[{{Tueller} {et~al.}(2010){Tueller}, {Baumgartner}, {Markwardt},
  {Skinner}, {Mushotzky}, {Ajello}, {Barthelmy}, {Beardmore}, {Brandt},
  {Burrows}, {Chincarini}, {Campana}, {Cummings}, {Cusumano}, {Evans},
  {Fenimore}, {Gehrels}, {Godet}, {Grupe}, {Holland}, {Kennea}, {Krimm},
  {Koss}, {Moretti}, {Mukai}, {Osborne}, {Okajima}, {Pagani}, {Page}, {Palmer},
  {Parsons}, {Schneider}, {Sakamoto}, {Sambruna}, {Sato}, {Stamatikos},
  {Stroh}, {Ukwata}, \& {Winter}}]{2010ApJS..186..378T}
{Tueller}, J. {et~al.} 2010, \apjs, 186, 378

\bibitem[{{Tueller} {et~al.}(2008){Tueller}, {Mushotzky}, {Barthelmy},
  {Cannizzo}, {Gehrels}, {Markwardt}, {Skinner}, \&
  {Winter}}]{2008ApJ...681..113T}
{Tueller}, J., {Mushotzky}, R.~F., {Barthelmy}, S., {Cannizzo}, J.~K.,
  {Gehrels}, N., {Markwardt}, C.~B., {Skinner}, G.~K., \& {Winter}, L.~M. 2008,
  \apj, 681, 113

\bibitem[{{Weaver} {et~al.}(2010){Weaver}, {Mel{\'e}ndez}, {Mushotzky},
  {Kraemer}, {Engle}, {Malumuth}, {Tueller}, {Markwardt}, {Berghea}, {Dudik},
  {Winter}, \& {Armus}}]{2010ApJ...716.1151W}
{Weaver}, K.~A. {et~al.} 2010, \apj, 716, 1151

\bibitem[{{Winter} {et~al.}(2010){Winter}, {Lewis}, {Koss}, {Veilleux},
  {Keeney}, \& {Mushotzky}}]{2010ApJ...710..503W}
{Winter}, L.~M., {Lewis}, K.~T., {Koss}, M., {Veilleux}, S., {Keeney}, B., \&
  {Mushotzky}, R.~F. 2010, \apj, 710, 503

\bibitem[{{Winter} {et~al.}(2009){Winter}, {Mushotzky}, {Reynolds}, \&
  {Tueller}}]{2009ApJ...690.1322W}
{Winter}, L.~M., {Mushotzky}, R.~F., {Reynolds}, C.~S., \& {Tueller}, J. 2009,
  \apj, 690, 1322

\bibitem[{{Winter} {et~al.}(2008){Winter}, {Mushotzky}, {Tueller}, \&
  {Markwardt}}]{2008ApJ...674..686W}
{Winter}, L.~M., {Mushotzky}, R.~F., {Tueller}, J., \& {Markwardt}, C. 2008,
  \apj, 674, 686

\end{thebibliography}

\begin{deluxetable}{lclllll}
\tablecaption{Warm Absorption through the Detection of O\,VII and O\,VIII Absorption Edges\label{tbl-warmabs}}
\tablewidth{0pt}
\tablehead{
\colhead{Source} & \colhead{Type} & \colhead{$\Delta\chi^2$} & \colhead{O\,VII} & \colhead{O\,VII} &  \colhead{O\,VIII} & \colhead{O\,VIII} \\
\colhead{} & \colhead{} & \colhead{} & \colhead{$E$ (keV)} & \colhead{$\tau$} & \colhead{$E$ (keV)} & \colhead{$\tau$} 
}
\startdata
\multicolumn{7}{c}{\bf Sources With Outflows Detected} \\
\hline
NGC 526A        & 1.5 & 17.99      & $0.70^{+0.01}_{-0.70}$    & $1.354^{+0.243}_{-0.256}$  & 0.87                      & $0.763^{+0.087}_{-0.099}$  \\
NGC 931         & 1.5 & 7008.86    & 0.73                      & $0.453^{+0.027}_{-0.029}$  & $0.87^{+0.01}_{-0.01}$    & $0.096^{+0.016}_{-0.019}$  \\
Mrk 6           & 1.5 & 94.69      & 0.73                      & $1.907^{+0.312}_{-0.289}$  & 0.87                      & $0.951^{+0.336}_{-0.625}$  \\
Mrk 79          & 1.2 & 134.82     & $0.72^{+0.01}_{-0.01}$    & $0.424^{+0.047}_{-0.044}$  & $0.90^{+0.01}_{-0.01}$    & $0.217^{+0.019}_{-0.042}$  \\
NGC 3227        & 1.5 & 757.7      & $0.72^{+0.01}_{-0.01}$    & $0.293^{+0.023}_{-0.024}$  & $0.85^{+0.02}_{-0.01}$    & $0.119^{+0.022}_{-0.023}$  \\
NGC 3516        & 1.5 & $-$988.44  & 0.73                      & 2.539                      & 0.87                      & 1.084                      \\
NGC 3783        & 1.5 &  6252.4     & $0.73^{+0.001}_{-0.001}$  & $1.037^{+0.025}_{-0.022}$  & $0.88^{+0.002}_{-0.002}$  & $0.660^{+0.018}_{-0.018}$  \\
NGC 4051        & 1.5 & 129.3      & 0.74                      & 0.072                      & 0.92                      & 0.031                      \\
NGC 4151        & 1.5 & 1790.61    & 0.73                      & 1.921                      & 0.87                      & 1.657                      \\
Mrk 766         & 1.5 & 537.33     & $0.73^{+0.01}_{-0.01}$    & $0.412^{+0.035}_{-0.031}$  & 0.87                      & $0.062^{+0.028}_{-0.024}$  \\
NGC 4593        & 1.0 &  78.4       & $0.73^{+0.02}_{-0.02}$    & $0.151^{+0.043}_{-0.046}$  & $0.87^{+0.03}_{-0.02}$    & $0.118^{+0.034}_{-0.034}$  \\
MCG -06-30-015  & 1.2 &  7333.1     & $0.71^{+0.001}_{-0.001}$  & $0.714^{+0.016}_{-0.017}$  & $0.86^{+0.003}_{-0.003}$  & $0.290^{+0.014}_{-0.013}$  \\
Mrk 290         & 1.0 & 146.9      & $0.72^{+0.02}_{-0.02}$    & $0.541^{+0.108}_{-0.116}$  & $0.87^{+0.03}_{-0.03}$    & $0.367^{+0.118}_{-0.107}$  \\
NGC 6860        & 1.5 &  94.3       & $0.73^{+0.02}_{-0.01}$    & $0.334^{+0.059}_{-0.093}$  & $0.87^{+0.01}_{-0.01}$    & $0.347^{+0.047}_{-0.055}$  \\
4C +74.26       & 1.0 & 61.2       & $0.73^{+0.01}_{-0.01}$    & $0.322^{+0.085}_{-0.024}$  & $0.91^{+0.01}_{-0.01}$    & $0.162^{+0.047}_{-0.029}$  \\
Mrk 509         & 1.2 & 28.3       & $0.73^{+0.03}_{-0.73}$    & $0.100^{+0.043}_{-0.060}$  & $0.85^{+0.02}_{-0.03}$    & $0.066^{+0.040}_{-0.046}$  \\
MR 2251-178     & 1.0 & 194        & 0.73                      & $0.276^{+0.029}_{-0.032}$  & 0.87                      & $0.000^{+0.010}_{-0.000}$  \\
NGC 7469        & 1.2 & 90.9       & $0.75^{+0.03}_{-0.03}$    & $0.074^{+0.029}_{-0.029}$  & $0.89^{+0.02}_{-0.02}$    & $0.089^{+0.024}_{-0.025}$  \\

\hline
\multicolumn{7}{c}{\bf Sources Without Outflows Detected}\\
\hline

Mrk 352                  & 1.0 & 4.5       & $0.71^{+0.02}_{-0.71}$  & $0.286^{+0.059}_{-0.236}$  & 0.87                    & $0.001^{+0.119}_{-0.001}$  \\
Fairall 9                & 1.0 &  12.1      & $0.71^{+0.02}_{-0.71}$  & $0.044^{+0.017}_{-0.020}$  & 0.87                    & $0.000^{+0.005}_{-0.000}$  \\
Mrk 1018                 & 1.5 & $-$0.02   & 0.73                    & $0.007^{+0.138}_{-0.007}$  & 0.87                    & $0.000^{+0.036}_{-0.000}$  \\
Mrk 590                  & 1.2 & $-$0.17   & 0.73                    & $0.000^{+0.030}_{-0.000}$  & 0.87                    & $0.000^{+0.012}_{-0.000}$  \\
ESO 198-024              & 1.0 & 7.89      & 0.73                    & $0.000^{+0.006}_{-0.000}$  & 0.87                    & $0.000^{+0.005}_{-0.000}$  \\
ESO 548-G081             & 1.0 & $-$0.05   & 0.73                    & $0.000^{+0.013}_{-0.000}$  & 0.87                    & $0.000^{+0.031}_{-0.000}$  \\
1H 0419-577              & 1.0 &  2.28      & 0.73                    & $0.000^{+0.003}_{-0.000}$  & 0.87                    & $0.000^{+0.001}_{-0.000}$  \\
3C 120                   & 1.0 & 12.53     & 0.73                    & $0.003^{+0.012}_{-0.003}$  & $0.90^{+0.03}_{-0.03}$  & $0.021^{+0.009}_{-0.009}$  \\
MCG -01-13-025           & 1.2 & 3.38      & 0.73                    & $0.271^{+0.132}_{-0.206}$  & 0.87                    & $0.000^{+0.034}_{-0.000}$  \\
Ark 120                  & 1.0 & 0.44      & 0.73                    & $0.000^{+0.005}_{-0.000}$  & 0.87                    & $0.000^{+0.002}_{-0.000}$  \\
ESO 362-G018             & 1.5 & 0.72      & 0.73                    & $0.010^{+0.031}_{-0.010}$  & 0.87                    & $0.000^{+0.013}_{-0.000}$  \\
PICTOR A                 & 1.0 & 1.8       & 0.73                    & $0.037^{+0.019}_{-0.037}$  & 0.87                    & $0.000^{+0.008}_{-0.000}$  \\
EXO 055620-3820.2        & 1.0 & $-$4.19   & 0.73                    & $0.000^{+0.047}_{-0.000}$  & 0.87                    & $0.000^{+0.012}_{-0.000}$  \\
2MASX J09043699+5536025  & 1.0 & 12.99     & 0.73                    & $0.267^{+0.204}_{-0.205}$  & $0.88^{+0.04}_{-0.04}$  & $0.259^{+0.141}_{-0.123}$  \\
MCG +04-22-042           & 1.2 & 0.21      & 0.74                    & $0.000^{+0.013}_{-0.000}$  & 0.87                    & $0.012^{+0.029}_{-0.012}$  \\
Mrk 110                  & 1.0 & 6.68      & 0.73                    & $0.124^{+0.036}_{-0.040}$  & 0.87                    & $0.000^{+0.011}_{-0.000}$  \\
UGC 06728                & 1.2 & 3.8       & 0.73                    & $0.023^{+0.058}_{-0.023}$  & 0.87                    & $0.017^{+0.037}_{-0.017}$  \\
IC 4329A                 & 1.2 & $-$11.69  & 0.73                    & $0.263^{+0.052}_{-0.054}$  & $0.86^{+0.01}_{-0.01}$  & $0.187^{+0.036}_{-0.023}$  \\
Mrk 279                  & 1.5 & 1.88      & 0.73                    & $0.014^{+0.016}_{-0.014}$  & 0.87                    & $0.000^{+0.004}_{-0.000}$  \\
NGC 5548                 & 1.5 & $-$1.96   & 0.73                    & $0.054^{+0.105}_{-0.054}$  & 0.87                    & $0.000^{+0.065}_{-0.000}$  \\
ESO 511-G030             & 1.0 & $-$0.29   & 0.73                    & $0.000^{+0.006}_{-0.000}$  & 0.87                    & $0.000^{+0.002}_{-0.000}$  \\
Mrk 841                  & 1.0 & 13        & $0.76^{+0.04}_{-0.03}$  & $0.067^{+0.045}_{-0.045}$  & 0.87                    & $0.000^{+0.030}_{-0.000}$  \\
3C 382                   & 1.0 & 7.46      & 0.73                    & $0.000^{+0.003}_{-0.000}$  & $0.88^{+0.03}_{-0.03}$  & $0.020^{+0.010}_{-0.010}$  \\
3C 390.3                 & 1.0 & $-$0.32   & 0.73                    & $0.000^{+0.007}_{-0.000}$  & 0.87                    & $0.000^{+0.008}_{-0.000}$  \\
NGC 7213                 & 1.5 & $-$59.3   & 0.73                    & $0.000^{+0.005}_{-0.000}$  & 0.87                    & $0.010^{+0.016}_{-0.010}$  \\
Mrk 926                  & 1.5 & $-$1.53   & 0.73                    & $0.000^{+0.025}_{-0.000}$  & 0.87                    & $0.000^{+0.022}_{-0.000}$  \\
\enddata
\vspace{0.2cm}
This table includes results of fitting {\tt zedge} models to the O\,VII and O\,VIII absorption edges at 0.73\,keV and 0.87\,keV, respectively.  Type is the optical Seyfert type for each AGN.  The $\Delta\chi^2$ measurement represents the improvement of the fit upon adding the edges to our baseline model.  We record both the measured edge energies (keV) and optical depths ($\tau$).  Where the energies were not well-constrained, we fixed these to the lab value.  For NGC 3516, NGC 4051, and NGC 4151, our baseline model was not sufficient to obtain a good fit to their spectra ($\chi^2/dof > 2$).  Therefore, error bars are not computed for these sources.
\end{deluxetable}

\begin{deluxetable}{l|ll|ll}
\tablecaption{Comparison of the Properties of Outflow vs. Non-outflow Sources\label{tbl-compare}}
\tablewidth{0pt}
\tablehead{
\colhead{Parameter} & \colhead{Outflow} & \colhead{Non-outflow} & \colhead{K-S $D$} & \colhead{K-S $P$} 
}
\startdata
$\log$ L$_{14-195\,{\rm keV}}$ & $43.38 \pm 0.79$ & $43.78 \pm 0.63$ & 0.5256 & 0.003\\
$\log$ L$_{14-195\,{\rm keV}}$/L$_{Edd}$ & $-2.92 \pm 0.49$ & $-2.80 \pm 0.64$ & 0.2265 & 0.588\\
$\log$ M/M$_{\sun}$ & $7.80 \pm 0.52$ & $8.09 \pm 0.63$ & 0.3419 & 0.131\\
$\log$ N$_{\rm H}$ & $20.57 \pm 1.12$ & $19.42 \pm 0.99$ & 0.4487 & 0.018\\
\enddata
\vspace{0.2cm}
This table includes the average and standard deviations on indicated parameters for the outflow and non-outflow sources.  Units for luminosity and N$_{\rm H}$ are ergs\,s$^{-1}$ and cm$^{-2}$, respectively.  Both the K-S D parameter and probability for the distributions being drawn from different populations are also given.
\end{deluxetable}

\begin{figure}
\centering
\includegraphics[width=8cm]{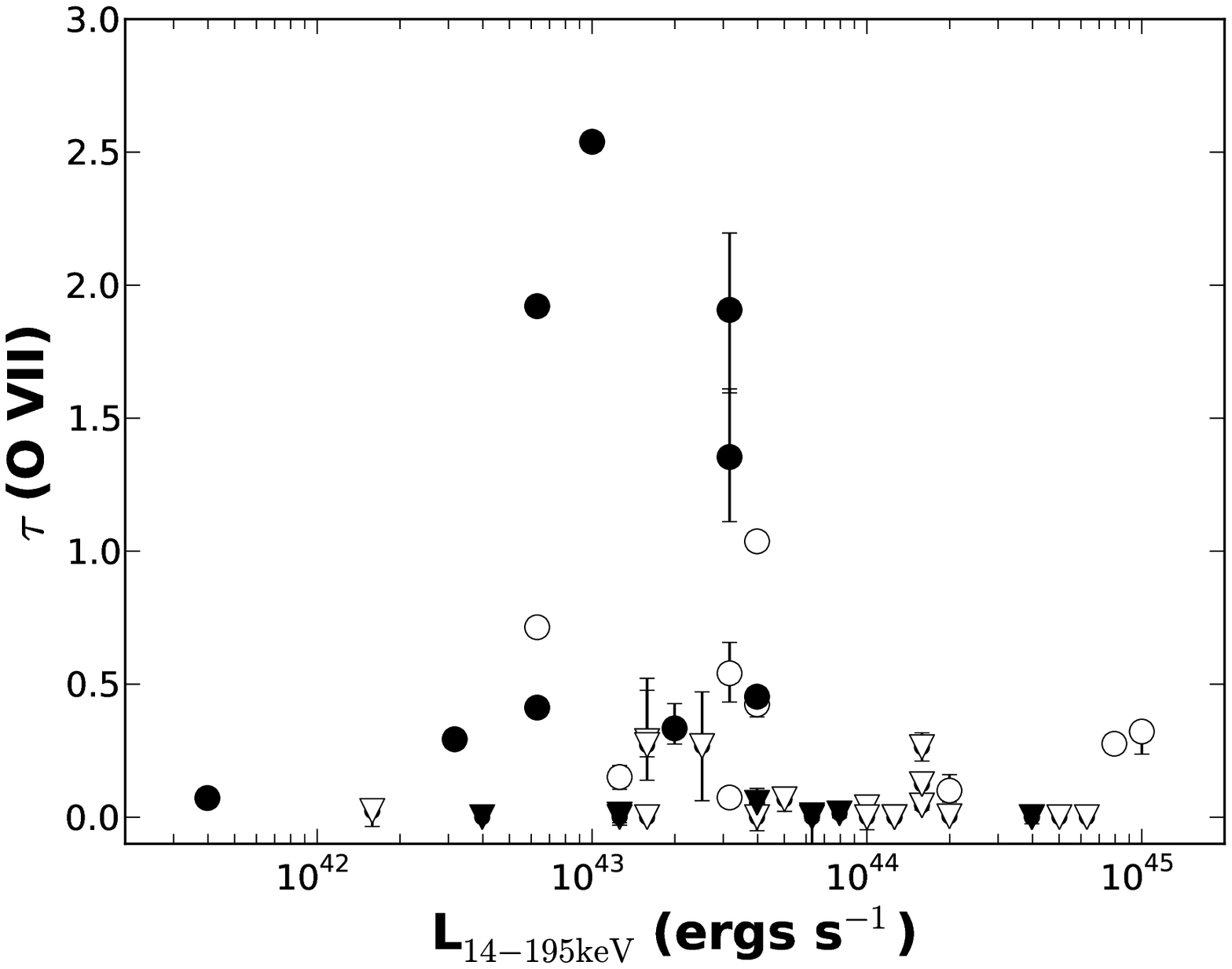}
\hspace{-0.5cm}
\includegraphics[width=8cm]{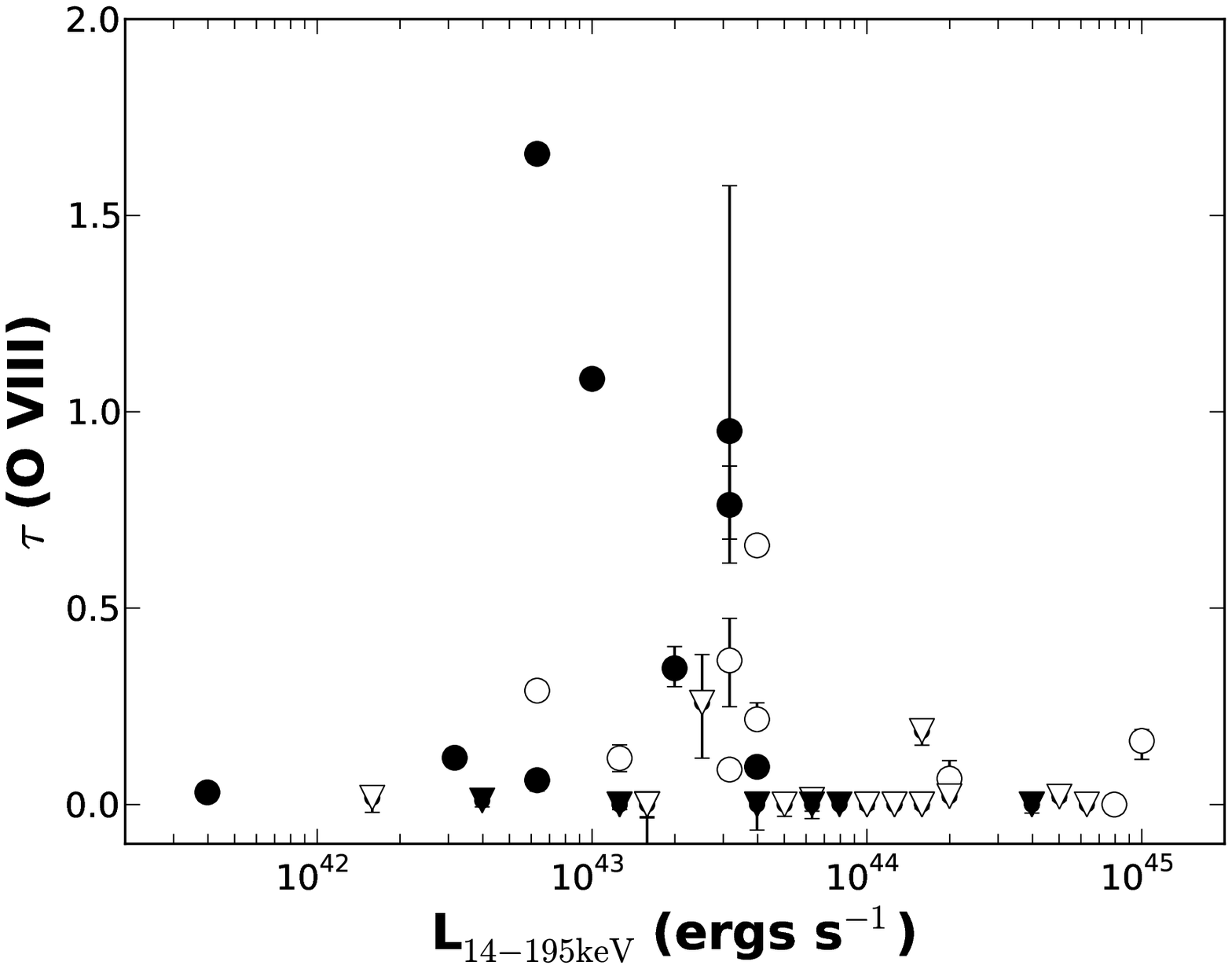} \\
\includegraphics[width=8cm]{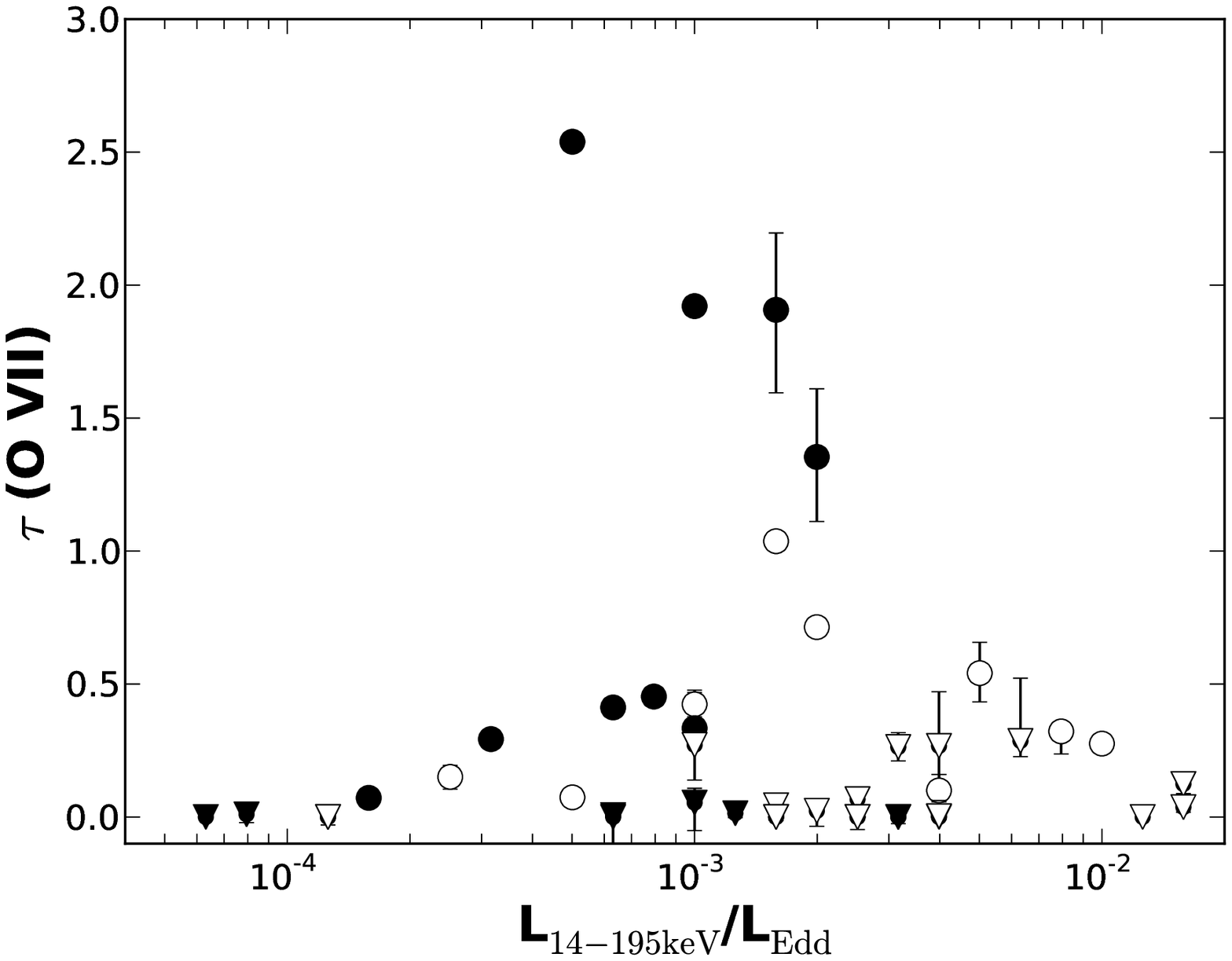}
\hspace{-0.5cm}
\includegraphics[width=8cm]{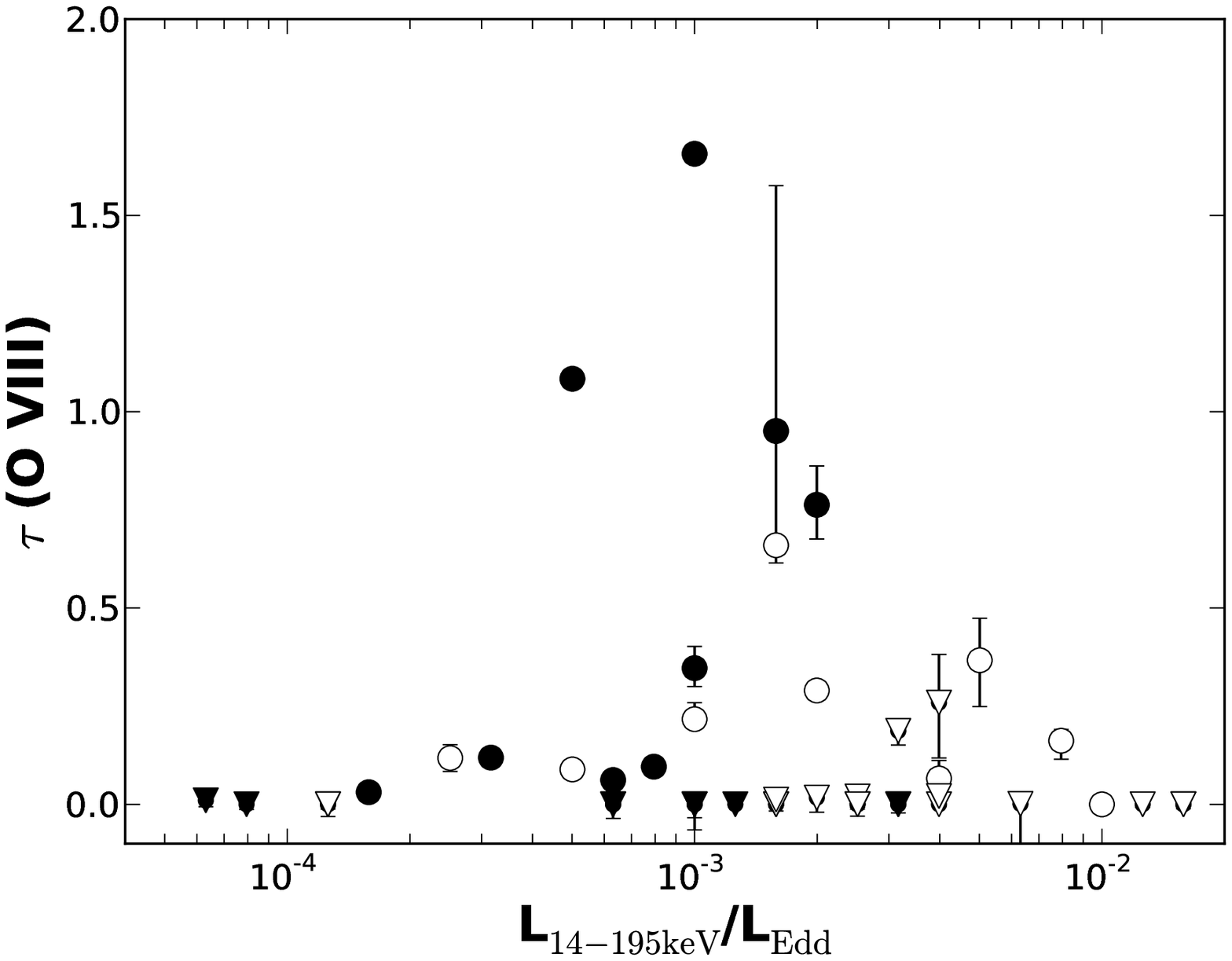}

\caption{Plotted is the optical depth of \ion{O}{7} (left) and \ion{O}{8} (right) versus the Swift BAT luminosity (top) and the ratio of the Swift BAT luminosity to the Eddington luminosity (bottom).  Sources where $\Delta\chi^2$ significantly improved upon adding the edge models (outflow sources) are shown with circles.  Non-outflow sources are represented by triangles.  Additionally, open symbols represent Sy 1--1.2s, while filled symbols represent Sy 1.5s.  The highest luminosity sources, as well as sources with the highest accretion rates, tend to be Sy 1--1.2s.  
}
\end{figure}

\begin{figure}
\centering
\includegraphics[width=8cm]{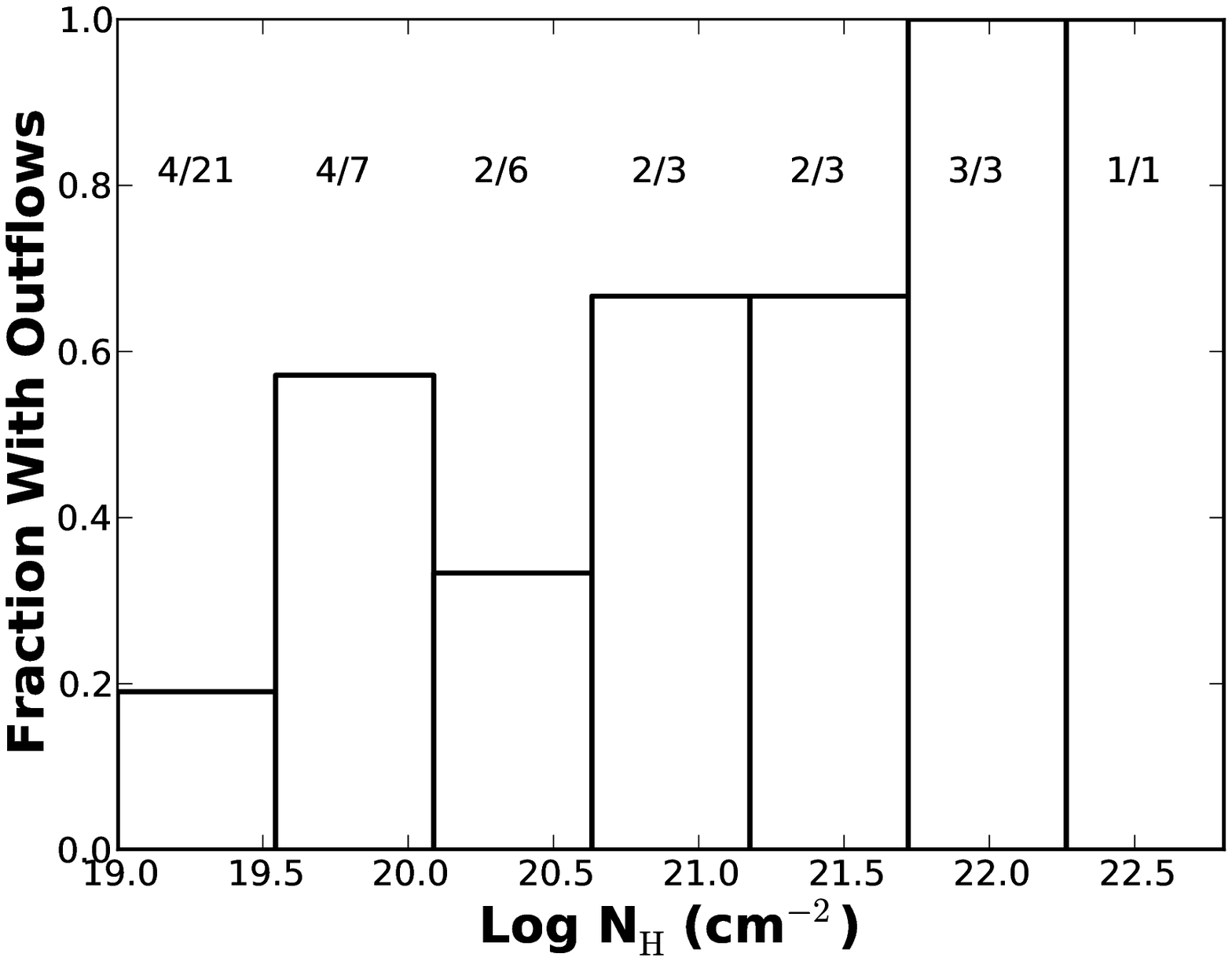}
\hspace{-0.5cm}
\includegraphics[width=8cm]{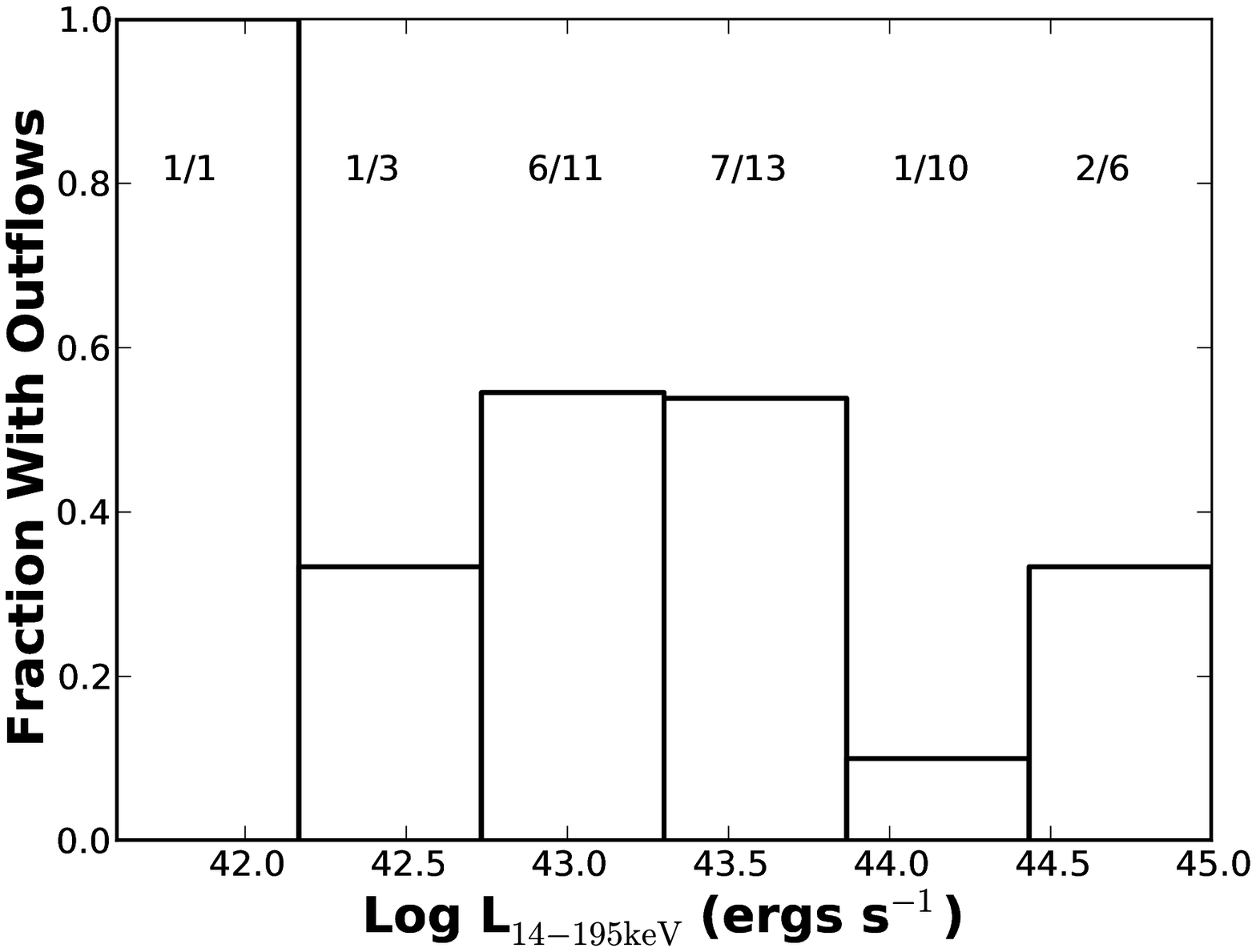}
\includegraphics[width=8cm]{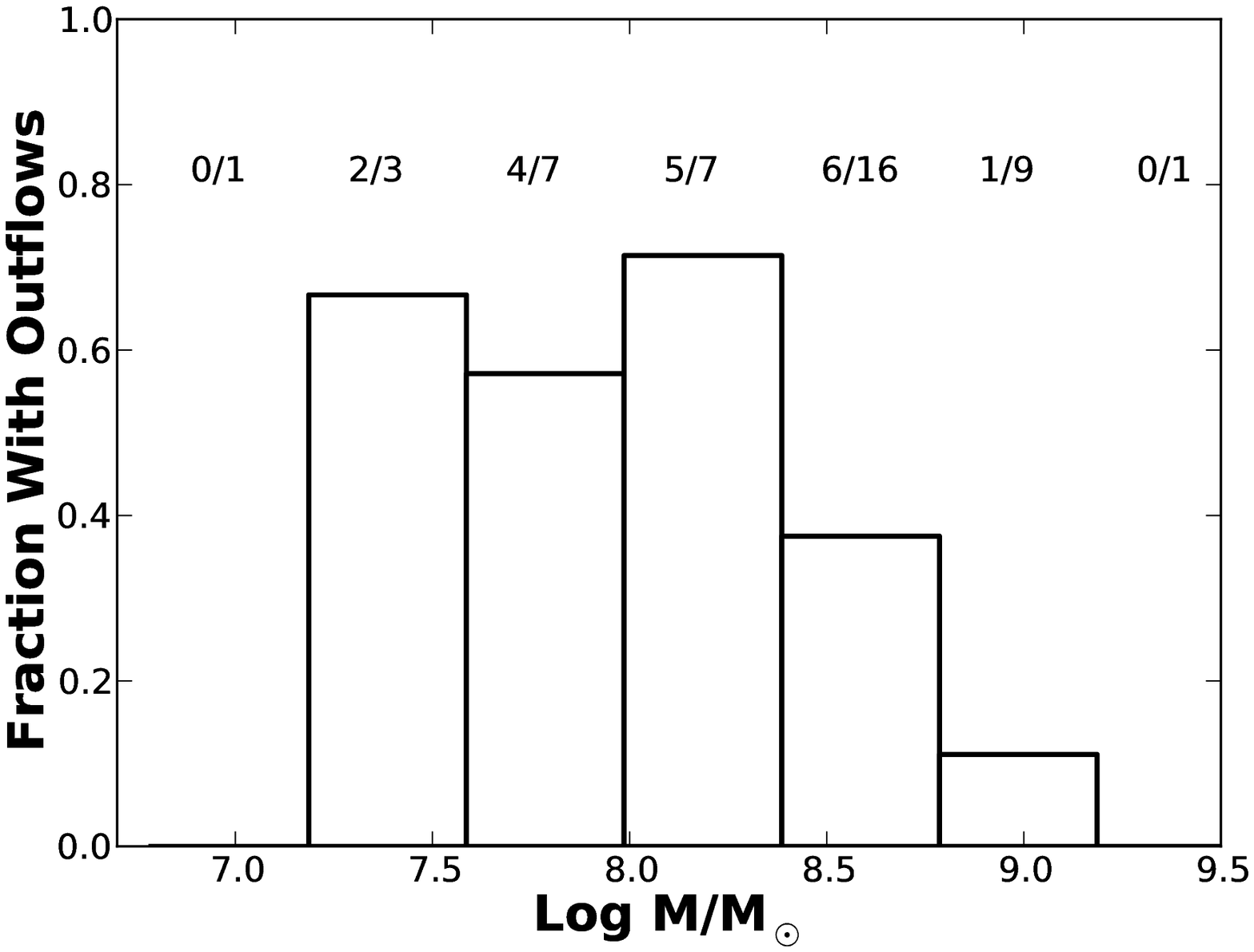}
\hspace{-0.5cm}
\includegraphics[width=8cm]{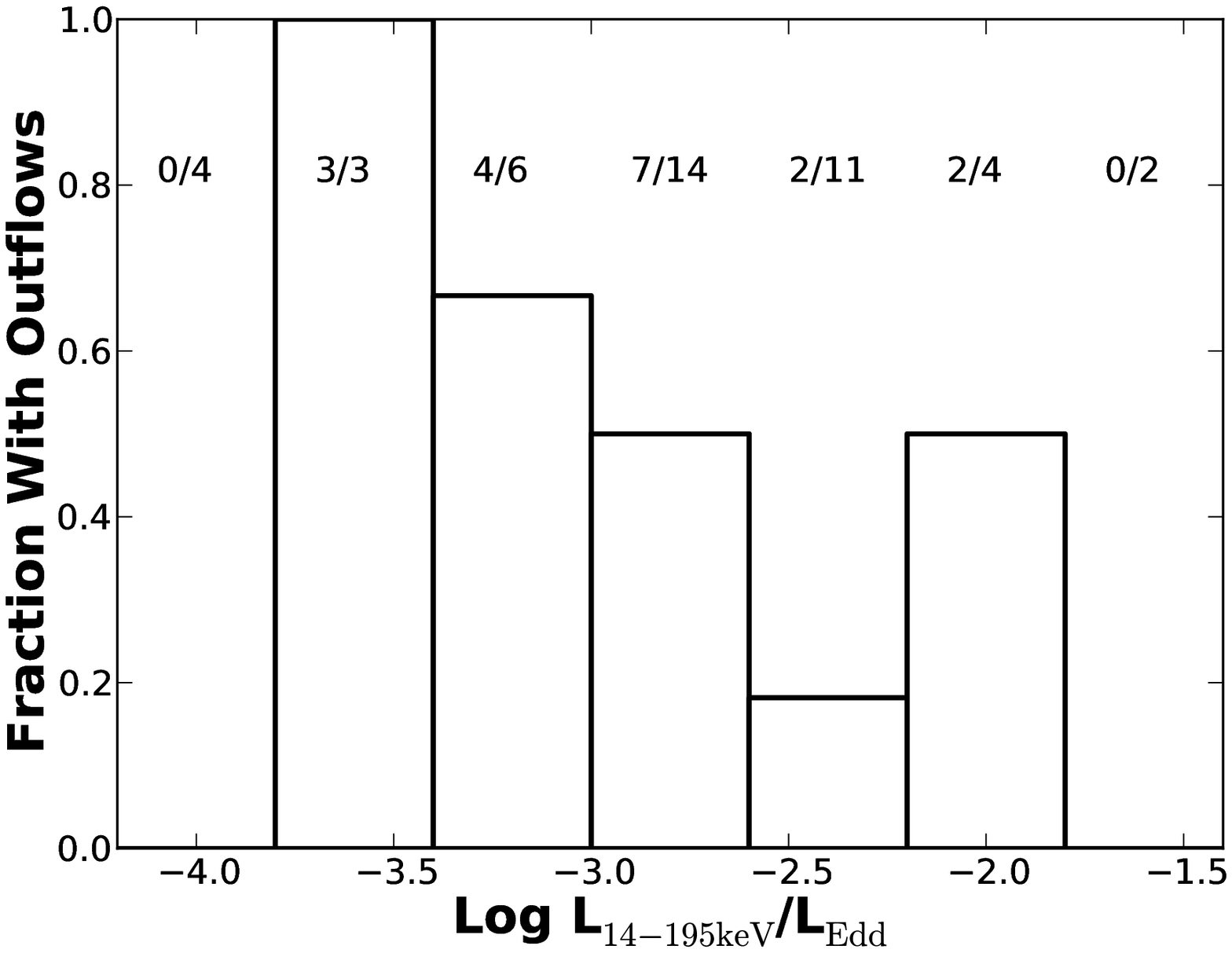}
\caption{Plotted, we show the fraction of outflow sources/total sources in binned intrinsic neutral column density (top left), Swift BAT luminosity (top right), black hole mass (bottom left), and accretion rate (bottom right).  We label the number of outflow sources out of the total number of sources in each bin.  Outflows tend to be detected in sources with higher N$_{\rm H}$.  There are few outflows detected in the most luminous sources, sources with the most massive black holes, and sources with the highest accretion rates.
}
\end{figure}

\end{document}